\documentclass[12pt]{article}
\usepackage{sc3conf}
\usepackage{amsfonts}
\usepackage{amssymb}
\usepackage{epsfig}

\usepackage{graphicx}

\begin{document}
\raggedbottom

\title{Internal structure of a thin transonic disk}
\authors{V.S.~Beskin\adref{1} and A.D.~Tchekhovskoy\adref{2}}
\addresses{\1ad P.N.Lebedev Physical Institute, Leninsky prosp., 53,
Moscow, 119991, Russia
\nextaddress \2ad Moscow Institute of Physics and Technology,
Dolgoprudny, Institutsky per., 9, 141700, Russia}

\maketitle

\begin{abstract}
The internal structure of the thin transonic disk accreting onto a
nonrotating black hole inside the last stable orbit ($r < 3r_{\rm g}$) is
considered within the hydrodynamical version of the Grad-Shafranov
equation. It is shown that in the vicinity of the sonic surface
takes place a sharp diminishing of the disk thickness. As a result,
in the vertical balance equation the dynamical forces
$\rho[({\bf v}\nabla){\bf v}]_{\theta}$ become important, these
on the sonic surface being of the same order as the pressure gradient
$\nabla_{\theta} P$. In the supersonic region
the thickness of the disk
is determined by the form of ballistic trajectories rather than by
the pressure gradient.
\end{abstract}

\section{Introduction}
According to the standard disk model~\cite{1},
the matter forms a thin balanced disk and
performs a circular motion with keplerian velocity
$v_{\rm K}(r) = \left(G{\cal M}/r\right)^{1/2}$.
The disk is thin provided that the accreting gas temperature is sufficiently
low ($c_{\rm s} \ll v_{\rm K}$), so that
$H \approx r c_{\rm s}/v_{\rm K}.$
Introducing the viscosity parameter $\alpha_{\rm SS} \leq 1$, relating the
stress tensor $t_{\varphi}^r$ and the pressure as
$t^r_{\varphi} = \alpha_{\rm SS}P$~\cite{1},
one can obtain
\begin{equation}
v_r/v_{ \rm K} \approx \alpha_{\rm SS} \, c_{\rm s}^2/v_{\rm K}^2.
\label{3}
\end{equation}

General relativity effects result in two important properties:
the absence of stable circular orbits at small radii
($r < 3r_{\rm g}$ for a nonrotating black hole,
\hbox{$r_{ \rm g} = 2G{\cal M}/c^2$} is the gravitational radius),
and the transonic regime of accretion.
The former means that for $r < 3r_{\rm g}$
a flow can be realized in the absence of viscosity.
The latter results from the fact that according to (\ref{3}) the flow is
subsonic outside the marginally stable orbit $r = 3 r_{\rm g}$ while at the
horizon $r = r_{\rm g}$ the flow is to be supersonic.

Up to now in the majority of works the procedure of vertical averaging
was used, where vertical four-velocity $u_{\hat\theta}$ was assumed to
be small~\cite{2,3}. As a result, the vertical component of the dynamic
force $nu^{b}\nabla_{b}(\mu u_{\hat\theta})$ was postulated to be
unimportant up to horizon. Here we are going to demonstrate that the
assumption $u_{\hat \theta} = 0$ is not correct. As in the Bondi
accretion, the dynamic force is to be important in the vicinity of the
sonic surface.

\section{Basic equations}

Consider ideal gas accreting onto a black hole
inside the marginally stable orbit $r = 3r_{\rm g}$.
Exact equations of motion of ideal media in Kerr metric were
formulated in~\cite{4}.
We consider a non-rotating (Schwarzschild) black hole only.

In Boyer-Lindquist coordinates ($t,r,\theta, \varphi$) the
Schwarzschild metric is~\cite{1}
${\rm d}s^{2}=-\alpha^{2}{\rm d}t^{2} + g_{ik}{\rm d}x^{i}{\rm d}x^{k},$
where
$\alpha^2 = 1-2{\cal M}/r, \, g_{rr}=\alpha^{-2}, \, g_{\theta \theta} = r^2$,
and $g_{\varphi \varphi} = \varpi^{2} = r^2\sin^2\theta.$
Here indices without caps denote vector components with
respect to the coordinate basis $\partial/\partial r$,
$\partial/\partial \theta$, $\partial/\partial \varphi$
and indices with
caps denote their physical components. The ${\bf\nabla}_k$ symbol
always represents a covariant derivative in space with $g_{ik}$ metric.
Finally, below we always use a system of units with $c=G=1$.

It is convenient
to introduce a stream function $\Phi(r,\theta)$. This function
defines the physical poloidal four-velocity component ${\bf u}_{\rm p}$
as
\begin{equation}
\alpha n{\bf u}_{\rm p}=
\frac{1}{2\pi\varpi}({\bf\nabla}\Phi\times{\bf e} _{\hat\varphi}).
\label{b2}
\end{equation}
Here $n$ is the particle concentration in the comoving reference
frame. It is the curves $\Phi(r,\theta)=$ const that define lines
of flow of the matter.

Three integrals of motion are conserved on $\Phi(r,\theta)= $
const surfaces: entropy $S = S(\Phi)$,
energy $E(\Phi)$, and $z$-component of angular momentum
$L(\Phi)$
\begin{equation}
E(\Phi) = \mu\alpha\gamma, \qquad
L(\Phi) = \mu\varpi u_{\hat\varphi}. \label{b3}
\end{equation}
Here $\mu=(\rho_{m}+P)/n$ ($\rho_{m}$ is internal energy density)
is the relativistic enthalpy. Below for
simplicity we use the polytropic equation of state
$P = k(S)n^{\Gamma}$,
so that the temperature and the sound velocity can be written as
\cite{1}
\begin{equation}
T = k(S)n^{\Gamma - 1}, \qquad c_{\rm s}^2 =
\frac{\Gamma}{\mu}k(S)n^{\Gamma - 1}. \label{Tc}
\end{equation}

As a result, the relativistic Euler equation
\begin{equation}
nu^{b}\nabla_{b}(\mu u_{a})+\nabla_{a}P - \mu n(u_{\hat
\varphi})^{2} \frac{1}{\varpi}\nabla_{a}\varpi+ \mu
n\gamma^{2}\frac{1}{\alpha}\nabla_{a}\alpha = 0
\label{euler}
\end{equation}
can be rewritten in the form of the Grad-Shafranov scalar equation
for the stream function
$\Phi(r,\theta)$ containing three integrals $E(\Phi)$,
$L(\Phi)$, and $S(\Phi)$~\cite{4}
\begin{eqnarray}
 & & -M^{2}\left[\frac{1}{\alpha}\nabla_{k}
\left(\frac{1}{\alpha\varpi
^{2}}\nabla^{k}\Phi\right)+\frac{1}{\alpha^{2}\varpi^{2}(\nabla
\Phi)^2}\frac{\nabla^{i}\Phi\nabla^{k}\Phi\nabla_{i}\nabla_{k}\Phi}{D}
\right]
\nonumber \\
\label{gsrel} \\
 & & +\frac{M^{2}\nabla'_{k}F\nabla^{k}\Phi}{2\alpha^{2}
\varpi^{2}({\bf\nabla}\Phi)^{2}D}
        +\frac{64\pi^{4}}{\alpha^{2}\varpi^{2}M^{2}}
\left(\varpi^{2}E\frac{{\rm d}E}{{\rm d}\Phi}
-\alpha^{2}L\frac{{\rm d}L}{{\rm d}\Phi}\right)
-16\pi^{3}nT\frac{{\rm d}S}{{\rm d}\Phi}=0, \nonumber
\end{eqnarray}
in which the $\nabla'_{k}$ derivative acts on all the variables
except the quantity $M^2$. Here the thermodynamical function $M^{2}$
is defined as $M^{2}=4\pi\mu/n$ and
\begin{equation}
D=-1+\frac{1}{u^{2}_{\rm p}}\frac{c^{2}_{s}}{1-c^{2}_{s}},
\quad
F=\frac{64\pi^{4}}{M^{4}}\left[\varpi^{2}E^{2}
-\alpha^{2}L^{2}-\varpi^{2}\alpha^{2}\mu^{2}\right].
\label{b10}
\end{equation}
To make the system complete we need to supply Grad-Shafranov
equation~(\ref{gsrel}) with the relativistic Bernoulli equation
$u_{\rm p}^2 = \gamma^2-u_{\hat\varphi}^2 - 1$; the latter with
the help of (\ref{b3}) can be rewritten as follows:
\begin{equation}
u_{\rm p}^2 = (E^2-\alpha^2L^2/\varpi^2
-\alpha^2\mu^2)/\alpha^2\mu^2.
\label{up2full}
\end{equation}

\section{Subsonic flow}

First of all, let us consider the subsonic region in the very vicinity
of the marginally stable orbit $r = r_0 = 3r_{\rm g}$ where the poloidal
velocity is much smaller than that of sound. Then equation
(\ref{gsrel}) can be significantly simplified by neglecting
the terms proportional to $D^{-1} \sim u_{\rm p}^2/c_{\rm s}^2$.
As a result, we have
\begin{equation}
-\frac{M^2}{\alpha}\nabla_{k}
\left(\frac{\nabla^{k}\Phi}{\alpha\varpi^{2}}\right)
+ \frac{64\pi^{4}}{\alpha^{2}\varpi^{2}M^{2}}
\left(\varpi^{2}E\frac{{\rm d}E}{{\rm d}\Phi}
-\alpha^{2}L\frac{{\rm d}L}{{\rm d}\Phi}\right)
-16\pi^{3}nT\frac{{\rm d}S}{{\rm d}\Phi}=0.
\label{main'}
\end{equation}
This equation describing the subsonic flow is elliptical.
Hence, it is necessary to specify five boundary conditions
on the surface of the last stable orbit $r_0 = 3r_{\rm g}$ where
$\alpha_0 = \alpha(r_0) = \sqrt{2/3}$,
$u_{\hat\varphi}(r_0) = 1/\sqrt{3}$, and $\gamma_0 = \gamma(r_0) =
\sqrt{4/3}$~\cite{1}. We consider below the
case where the radial velocity is constant on the surface $r = r_0$
and the toroidal four-velocity is exactly equal to $u_{\hat\varphi}(r_0)$:
\begin{equation}
u_{\hat r}(r_0,\Theta)  =  -u_0, \quad
u_{\hat \Theta}(r_0,\Theta)   =  \Theta u_0, \quad
u_{\hat\varphi}(r_0,\Theta)  =  1/\sqrt{3}.
\label{new}
\end{equation}
Here $u_{\hat \Theta} \left(r_0,\Theta\right)$ corresponds to the plane flow
at the marginally stable orbit,
and we introduced the new angular variable $\Theta = \pi/2 -\theta$
($\Theta_{\rm disk} \sim c_0$) which is counted off from the equator in the
vertical direction. Next, we suppose that the velocity of sound is also a
constant on the surface $r = r_0$:
\begin{equation}
c_{\rm s}(r_0,\Theta) = c_0.
\label{c0}
\end{equation}
For $P = k(S)n^{\Gamma}$ this means that both the
temperature $T_0 =T(r_0)$ and the relativistic enthalpy $\mu_0 = \mu(r_0)$
are also constant on this surface. It is necessary to stress that, according
to (\ref{3}), for nonrelativistic temperature of the accreting gas $c_0 \ll 1$
we have a small parameter $u_0/c_0 \sim \alpha_{\rm SS}c_0 \ll 1$. Finally, as
the last, fifth, boundary condition it is convenient to specify the entropy
$S(\Phi)$.

Introducing now the values
$e_0 =\alpha_0\gamma_0 = \sqrt{8/9}$ and $l_0 = u_{\hat\varphi}(r_0)r_0
= \sqrt{3}r_{\rm g}$, one can rewrite the invariants $E(\Phi)$ and $L(\Phi)$
as
\begin{equation}
E(\Phi)  =  \mu_0 e_0 = {\rm const}, \qquad L(\Phi) = \mu_0 l_0 \cos\Theta_m.
\label{invval}
\end{equation}
Here $\Theta_m = \Theta_m(\Phi)$ is the angle for which $\Phi(r_0,
\Theta_m) = \Phi (r, \Theta)$. In other words, the function
$\Theta_m(r, \Theta)$ has the meaning of a theta angle on the last
stable orbit connected with a given point
($r$,$\Theta$) by a line of flow $\Phi(r,\Theta) = $ const. In
particular, $\Theta_m(r_0, \Theta) = \Theta$.

First of all, we see that condition $E = $ const (\ref{invval}) allows us
to rewrite equation (\ref{main'}) in a simpler form
\begin{equation}
\frac{\partial^2 \Phi}{\partial r^2}
+\frac{\cos\Theta}{\alpha^2r^2}\frac{\partial}{\partial\Theta}
\left(\frac{1}{\cos\Theta}\frac{\partial\Phi}{\partial
\Theta}\right)
= -4\pi^2n^2\frac{L}{\mu^2}\,\frac{{\rm d}L}{{\rm d}\Phi}
- 4\pi^{2}n^2r^2\cos^2\Theta\frac{T}{\mu}\,\frac{{\rm d}S}{{\rm d}\Phi}.
\label{main}
\end{equation}
Next, as one can show~\cite{6}, for $r=r_0$, the r.h.s.\hbox{} of
equation (\ref{main}) describes the transverse balance of a
pressure gradient and a gravitational force, whereas the
l.h.s.\hbox{} corresponds to the dynamic term $({\bf v}\nabla){\bf
v}$. At the marginally stable orbit it is of the order of
$u_0^2/c_0^2$ and may be dropped. It is therefore natural to
choose the entropy $S(\Phi)$ from the condition of a transverse
balance on the surface $r = r_0$
\begin{equation}
r_0^2\cos^2\Theta_m\frac{{\rm d}S}{{\rm d}\Theta_m} =
-\frac{\Gamma}{c_0^2}\,\frac{L}{\mu_0^2}\,\frac{{\rm d}L}{{\rm d}\Theta_m},
\label{sr0}
\end{equation}
where $L(\Theta_m)$ is determined from the boundary
condition (\ref{invval}). Thus, we have
\begin{equation}
S(\Theta_m) = S(0) - \frac{\Gamma}{3c_ 0^2}\ln(\cos\Theta_m).
\label{sp}
\end{equation}
Owing to (\ref{c0}), relation (\ref{sp}) corresponds to the standard
concentration profile
\begin{equation}
n(r_0, \Theta) \approx n_0
\exp\left(-\frac{\Gamma}{6c_0^2}\Theta^2\right).
\label{np}
\end{equation}
Finally, definition (\ref{b2}) results in the following relationship
between functions $\Phi$ and $\Theta_m$:
\begin{equation}
{\rm d}\Phi = 2\pi \alpha_0 r_0^{2}n(r_0, \Theta_m)u_0
\cos\Theta_m{\rm d}\Theta_m.
\label{d3}
\end{equation}
Hence, due to (\ref{invval}), (\ref{np}), and (\ref{d3}),
the invariant $L(\Phi)$ can be directly determined from the boundary
conditions as well.

Equation (\ref{main}) together with boundary conditions (\ref{new}),
(\ref{invval}), (\ref{sp}), (\ref{np}), and (\ref{d3}) determines
structure of the inviscid subsonic flow inside the marginally
stable orbit. For example, for $c_{\rm s} \ll 1$ we obtain using
(\ref{up2full})
\begin{equation}
u_{\rm p}^2  = u_0^2 +w^2 +
\frac{1}{3}\left(\Theta_m^2-\Theta^2\right) +\frac{2}{\Gamma -
1}(c_0^2 - c_{\rm s}^2) + \dots
\label{uup}
\end{equation}
Here the quantity $w$, where
\begin{equation}
w^2(r) = \frac{e_0^2-\alpha^2l_0^2/r^2-\alpha^2}{\alpha^2}
\approx \frac{1}{6}\, \left(\frac{r_0 - r}{r_0}\right)^3,
\label{w2}
\end{equation}
depending on the radius $r$ only, is a poloidal four-velocity of a
free particle having zero poloidal velocity for $r = r_0$. As we
see, $w^2$ increases very slowly when moving away from the last
stable orbit. Therefore, the contribution of $w^2$ turns out to be
negligibly small in the subsonic region.

\begin{figure}[htb]
  \begin{center}
    \includegraphics[bb = 143 458 471 817,
      height=7cm, keepaspectratio]{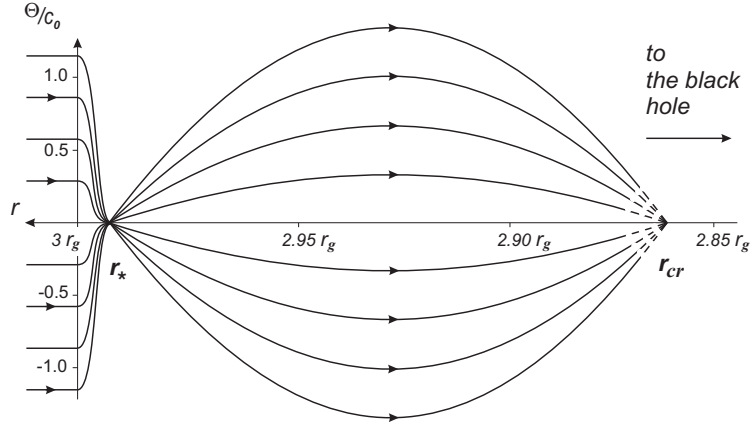}
         \caption{The structure of a thin accretion
          disk (actual scale) for $c_0 = 10^{-2}$, $u_0 = 10^{-5}$
          after passing the marginally stable orbit $r = 3r_{\rm g}$.
         In the vicinity of the sonic surface $r = r_*$ the flow
         has a form of an ordinary nozzle.}
     \label{figure}
   \end{center}
 \end{figure}

An important conclusion can be drawn directly from (\ref{uup}) in
which for the equatorial plane we have $\Theta_m = \Theta = 0.$
Assuming $u_{\rm p} = c_{\rm s} = c_*$ and neglecting $w^2$, we
find the velocity of sound $c_*$ on the
sonic surface $r = r_*$, $\Theta = 0$:
\begin{equation}
c_* \approx \sqrt{\frac{2}{\Gamma+1}}\,c_0.
\label{cc}
\end{equation}
As we see, $c_* \approx c_0$. Next, as the entropy $S$ remains constant along
the flow lines, the gas concentration remains approximately constant along
the flow lines ($n(r_*,\Theta) \approx n(r_0,\Theta_m)$) as well. In other words,
in agreement with the Bondi accretion, the subsonic flow can be considered
incompressible. On the other hand, because the radial velocity increases
from $u_0$ to $c_* \approx c_0$, i.e., for $c_0 \ll 1$ ($u_0/c_0 \ll 1$)
it changes over several orders of magnitude, the disk thickness $H$
should change in the same proportion owing to the continuity
equation (see Fig. \ref{figure})
\begin{equation}
H(r_*) \approx \frac{u_0}{c_0}H(r_0).
\label{compr}
\end{equation}
As a result, a rapid decrease of the disk thickness should be
accompanied by the appearance of vertical component of
velocity which also should be taken into account in Euler equation
(\ref{euler}).

Indeed, as one can find analyzing asymptotic of equation
(\ref{main})~\cite{6}, in the vicinity of the sonic surface
located at $r_*  =  r_0 - \Lambda u_0^{2/3}r_0$, where the
logarithmic factor
 $\Lambda = (3/2)^{2/3}[\ln(c_0/u_0)]^{2/3} \approx 5 - 7$,
the components of the velocity and the pressure gradient
in the limit $r \rightarrow r_*$ can be presented as
\begin{equation}
u_{\hat \Theta}  \rightarrow  - \frac{c_0}{u_0}\Theta, \quad
u_{\hat r}  \rightarrow  - c_{*},  \quad
- \frac{\nabla_{\hat \Theta}P}{\mu}  \rightarrow
\frac{c_0^2}{u_0^2}\, \frac{\Theta}{r}.
\end{equation}
On the other hand, near the sonic surface the radial scale $\delta
r$ determining the radial derivatives becomes as small as the
transverse dimension of a disk: $\delta r \approx H(r_*) \approx
u_0r_0$. Hence, logarithmic derivative
$\eta_1 = (r/n)(\partial n/\partial r)$
can be evaluated as $\eta_1 \approx u_0^{-1}$.
As a result, both components of the dynamic force
\begin{equation}
\frac{u_{\hat \Theta}}{r}\,\frac{\partial u_{\hat \Theta}}{\partial \Theta}
\rightarrow \frac{c_0^2}{u_0^2}\,\frac{\Theta}{r}, \quad
u_{\hat r}\frac{\partial u_{\hat \Theta}}{\partial r}  \rightarrow
\frac{c_0^2}{u_0^2}\,\frac{\Theta}{r},
\label{w6}
\end{equation}
do become of the order of the pressure gradient.

\section{Transonic flow}

To check our conclusion one can consider flow structure in the
vicinity of the sonic surface in more detail. Because the smooth transonic
flow is analytical at a singular point, one can write 
\begin{eqnarray}
 n & = & n_*\left(1+\eta_1h+
     \frac{1}{2}\eta_3\Theta^2+\dots\right),
\label{expansionn}\\
 \Theta_m &= & a_0\left(\Theta+a_1h\Theta+\frac{1}{2}a_2h^2\Theta+
            \frac{1}{6}b_0\Theta^3 + \dots
            \right),
\label{expansionthetam}
\end{eqnarray}
where $h = (r-r_*)/r_*$. Here we assume that all the three
invariants $E$, $L$, and $S$ are already given. Hence, the problem
needs one extra boundary condition. Now comparing the appropriate
coefficients in Bernoulli (\ref{up2full}) and full stream equation
(\ref{gsrel}), one can obtain neglecting terms $\sim u_0^2/c_0^2$
\begin{eqnarray}
a_0 & = & \left(\frac{2}{\Gamma+1}\right)^%
        {(\Gamma+1)/2(\Gamma-1)}\frac{c_0}{u_0},
\label{a_0}\\
a_1 & = & 2 + \frac{1-\alpha_*^2}{2\alpha_*^2} \approx 2.25,
\label{a_1} \\
a_2 & = & -(\Gamma+1)\eta_1^2,
\label{a_2}\\
b_0 & = &
 \left(\frac{\Gamma + 1}{6}\right) \frac{a_0^2}{c_0^2}, \\
\eta_3 & = & -\frac{2}{3} (\Gamma + 1)\eta_1^2
- \left(\frac{\Gamma-1}{3}\right)\frac{a_0^2}{c_0^2},
  \label{eta3}
\end{eqnarray}
where $\alpha_*^2 = \alpha^2(r_*) \approx 2/3$.

As we see, coefficients (\ref{a_0})--(\ref{eta3}) are expressed
through the radial logarithmic derivative $\eta_1$. They have clear
physical meaning. So, $a_0$ gives the compression of flow lines:
$a_0 = H(r_0)/H(r_*)$. In agreement with (\ref{compr}) we have
$a_0 \approx c_0/u_0$. Further, $a_1$ corresponds to the slope of
the flow lines with respect to the equatorial plane. As $a_1 > 0$,
the compression of stream line finishes somewhere before the sonic surface,
so inside the sonic radius $r < r_*$ the stream
lines diverge. On the other hand, as $a_1 \ll u_0^{-1}$, for $r =
r_{*}$ the divergency is still very weak. Hence, in the vicinity
of the sonic surface the flow has a form of an ordinary nozzle
(see Fig. 1). Finally, as $a_2 \sim \eta_3 \sim b_0 \sim
u_0^{-2}$, one can conclude that the transverse scale of the
transonic region $H(r_*)$ is the same as the longitudinal one. The
latter point suggests a very important consequence that the
transonic region is essentially two-dimensional, and it is
impossible to analyze it within the standard one-dimensional
approximation.

Let us stress that it is rather difficult to connect the sonic characteristics
$\eta_1 = \eta_1(r_*)$ with physical boundary conditions on the marginally
stable orbit $r = r_0$ (for this it is necessary to know all the expansion
coefficients in (\ref{expansionn}) and (\ref{expansionthetam})).
In particular, it is impossible to formulate the restriction on five
boundary conditions (\ref{new}), (\ref{c0}) and (\ref{sp}) resulting from
the critical condition on the sonic surface. Nevertheless,
the estimate $\eta_1 \approx u_0^{-1}$
makes us sure that we know the parameter $\eta_1$
to a high enough accuracy. Then, according to (\ref{a_0})--(\ref{eta3}), all
the other coefficients can be determined exactly.

Using now expansions (\ref{expansionn}) and (\ref{expansionthetam}), one can
obtain all other physical parameters of the transonic flow. In
particular, we have
\begin{eqnarray}
 u_{\rm p}^2 & = & c_*^2
   \left[1-2\eta_1 h+
         \frac{1}{6}(\Gamma - 1) \, \frac{a_0^2}{c_0^2}\Theta^2
                  + \frac{2}{3}(\Gamma + 1)\eta_1^2\Theta^2
   \right],
\nonumber  \\
 c_{\rm s}^2 & = & c_*^2
   \left[1+\left(\Gamma-1\right)\eta_1 h +
        \frac{1}{6}(\Gamma -1 ) \, \frac{a_0^2}{c_0^2}\Theta^2
       - \frac{1}{3}(\Gamma - 1)(\Gamma+1) \eta_1^2
         \Theta^2
   \right].
\nonumber
\end{eqnarray}
Hence, a shape of the sonic surface $u_{\rm p} = c_{\rm s}$
has the standard parabolic form
\begin{equation}h =
\frac{1}{3}\eta_1\Theta^2.
\end{equation}

\section{Supersonic flow}

Because the pressure gradient becomes insignificant in supersonic region,
the matter moves here along the trajectories of free
particles.
Neglecting $\nabla_{\theta}P$ term in the $\theta$-component of relativistic
Euler equation~(\ref{euler}), we have (cf.~\cite{3})
\begin{equation}
 \alpha u_{\hat r} \frac{\partial (r u_{\hat \Theta})} {\partial r}
 + {(r u_{\hat \Theta})\over r^2}
 \frac{\partial (r u_{\hat \Theta})} {\partial \Theta} +
 (u_{\hat\varphi})^2 \tan\Theta = 0.
\label{theuler}
\end{equation}
Here $u_{\hat\varphi}$ can be easily
expressed in terms of radius: $u_{\hat\varphi} = \sqrt{3}/x,$
where \hbox{$x = r/r_{\rm g}$}.
We also introduce dimensionless functions $f(x)$ and $g(x)$:
$\Theta f(x) = x u_{\hat \Theta}$,
$g(x) = -\alpha u_{\hat r} > 0$.
Using now (\ref{theuler}) and
definitions above, we obtain an ordinary differential equation for $f(x)$
\begin{equation}
\frac{{\rm d} f}{{\rm d} x} = \frac{f^2 + 3}{x^2 g(x)}.
\label{f}
\end{equation}
Next, analyzing equation (\ref{up2full}), one can conclude that
$u_{\rm p} \rightarrow w$ as $r \rightarrow r_{\rm g}$.
On the other hand, $u_{\rm p} \approx c_* \approx c_0$ for $r \lesssim r_*$.
Therefore, the following
approximation should be valid throughout $r_{\rm g} < r < r_*$ region:
$g(x) \approx \sqrt{(\alpha w)^2 + (\alpha c_0)^2}$,
where, owing to (\ref{w2}), $\left(\alpha w\right)^2 = (3-x)^3/(9x^3)$.

The results of calculations are presented on Fig.\hbox{}~\ref{figure}.
As we see, in the supersonic region the flow performs transversal
oscillation about the equatorial plane, their frequency independent of
the amplitude. Once diverged, the flow converges once again
at the point $r = r_{\rm cr}$ ($r_0 - r_{\rm cr} = Ac_*r_0$,
$A \approx 2\pi$) on the equatorial plane.
Such oscillation can be easily understood. Indeed, as one can see
from (\ref{uup}), for $c_0 \ll 1$ poloidal motion remains
nonrelativistic for $r \sim r_*$. Hence,
it is possible to use the nonrelativistic equation
\begin{equation}
\frac{{\rm d} v_{\hat\Theta}}{{\rm d} t} = - \Theta \frac{v_{\hat\varphi}^2}{r},
\label{f1}
\end{equation}
where $v_{\hat\Theta} = r{\rm d}\Theta/{\rm d} t$.
Because the spatial amplitude of oscillations $\sim c_0r_0$
is small compared with $r_0$,
one can write $v_{\hat\varphi} = u_{\hat\varphi}(r_0)/\gamma_0= 1/2$
(see (\ref{new})), $r = 3r_{\rm g}$, and $|v_r| \approx c_*$.
This gives $\Theta(h) = \Theta_{\rm A} \sin[(v_{\hat \varphi}/c_*)h]$,
$h = (r - r_*)/r_*$
in full agreement with exact calculations.
Clearly, additional consideration is necessary to determine
the flow structure for $r < r_{\rm cr}$. Nevertheless, one can be sure
that the accretion disk thickness, though oscillating in the supersonic
region, remains as small as in the region of stable orbits:
$\Theta_{\rm disk} \lesssim c_0$.

\section{Conclusion}

As was shown, the diminishing disk thickness in the vicinity of the
sonic surface inevitably leads to the
emergence of the vertical velocity component of the accreting matter.
As a result, the dynamic term $({\bf v}\nabla){\bf v}$ in the vertical
balance equation cannot be omitted.
It is necessary to stress that whereas at the sonic surface
both components of dynamic term
$[({\bf v}\nabla){\bf v}]_{\theta}$ (\ref{w6})
become of the same order of magnitude as the pressure gradient,
the role of the gravitational term remains unimportant:
$\nabla_{\theta}\varphi_{\rm g} \sim \Theta/r$,
i.e., it is $c_0^2/u_0^2$ times smaller than the leading terms.
As a result, the structure of a thin transonic disk is quite similar
to that of an ordinary planar nozzle.
For this reason the critical condition
on the sonic surface does not restrict the accretion rate.
We suppose that for a given accretion rate it determines
vertical component of the velocity on the marginally stable orbit
which does not affect the flow structure.
Finally, inside the sonic surface $r < r_*$ the pressure term
$\nabla_{\theta}P$ becomes unimportant, so the thickness of the disk
is determined by the form of ballistic trajectories.

This work was partially supported
by the Russian Foundation for Basic Research (grant No.~00--15--96594).

\end{document}